\documentstyle[prl,aps,tighten,multicol,epsfig]{revtex}

\begin{document}

\title{Out of Equilibrium Dynamics of the Hopfield Model in its spin--glass phase}

\author{Marcelo A. Montemurro$^1$, Francisco A. Tamarit$^1$\thanks{Member
of the National Research Council, CONICET (Argentina)}, Daniel A. Stariolo$^2$ and Sergio A. Cannas$^{1*}$}

\address{$^1$ Facultad de Matem\'atica, Astronom\'{\i}a y F\'{\i}sica, Universidad Nacional de C\'ordoba \\
Ciudad Universitaria, 5000 C\'ordoba, Argentina\\
$^2$ Instituto de F\'{\i}sica, Universidade Federal do Rio Grande do Sul \\
CP 15051, 91501-970 Porto Alegre, Brazil}

\maketitle

\begin{abstract}

In this work we study numerically the out of equilibrium dynamics of the Hopfield model for associative memory
inside its spin--glass phase. 
Besides its interest as a neural network 
model it can also be considered as a prototype of fully connected magnetic 
systems with randomness and frustration. By adjusting the ratio between the 
number of stored configurations $p$ and the total number of neurons $N$ one can control 
the phase--space structure, whose complexity can vary between the simple mean--field ferromagnet  (when $p=1$)
and that of the Sherrington--Kirkpatrick spin--glass model (for a properly taken limit of an infinite number of
patterns). In particular, little attention has been devoted to the spin--glass phase of this model. 
In this work we analyse the two--time 
auto--correlation function, the decay of the 
magnetization and the distribution of overlaps between states. The results show that within the 
spin--glass phase of the model the dynamics exhibits ageing phenomena and presents features that suggest a non 
trivial breaking of replica symmetry.
{\bf Pacs Numbers:} 75.10Nr;64.60Ht \\
\end{abstract}

\begin{multicols}{2}

\narrowtext

In recent years the off--equilibrium dynamics of spin--glasses below the freezing
temperature has been the subject of a great number of studies in the field of 
complex magnetic systems \cite{cugliandolo1}, both experimental and theoretical.
Real spin--glasses are characterized by such an extremely slow dynamics that they
may never attain equilibrium within experimental time scales. Under these
circumstances a theoretical description of the actual physics of real spin--glasses
requires a dynamical approach.

Statistical physics models, despite their simplifications, have shown to be 
very useful in the understanding of the behaviour of these materials.
Among them, perhaps the most relevant one in the development of this
subject, is the well known Edwards--Anderson model.
Since a complete analytical description of this model has not be achieved
up to now (due to the enormous mathematical difficulties involved), numerical simulations
emerged as the main tool of research in this area. However, in order to implement them, it is
necessary to provide the system with an adequate dynamics,
usually accomplished by means of a stochastic (Monte Carlo) process. 
Although these ad--doc dynamics
were originally introduced to compute equilibrium quantities, 
quite surprisingly they have also proved to be very useful in simulating
the actual dynamical processes observed in real materials \cite{zheng}. 
This agreement opened up a whole new range of possibilities in  statistical physics
research by allowing physicists to simulate, with simple models and
Monte Carlo dynamics, the complex out of equilibrium behaviour of spin--glasses
and other magnetic materials.

Concerning equilibrium properties, the  long range version of the EA model due to 
Sherington and Kirkpatrick \cite{sherrington} (SK model), 
has raised particular interest owing to the fact that 
an exact solution is known for its 
thermostatics \cite{parisi80} and insightful approximations 
have been found even for its 
dynamics \cite{cugliandolo2}. The picture that emerges for 
this model is that of a phase space with a very 
intricate hierarchical structure of basins, whose number and depth diverge with 
the size of the system---as depicted by the solution due to G. Parisi \cite{parisi80}.

 Moreover, within each of those basins, which divide the 
system in independent ergodic components, there is as well a complex structure of 
subvalleys within subvalleys separated by barriers with a continuous 
distribution of heights. Within the framework of this complicated phase--space 
geometry the ensuing dynamics turns out to be of an extremely slow character and a 
wide range of new time dependent phenomena are observed, which collectively are 
referred to as {\em ageing phenomena}. Starting from random 
initial conditions  such a system may never achieve true equilibrium; 
therefore, its inherent physics must be interpreted as dynamical in nature.
Whether this picture is shared by low dimensional systems or not, 
is still under enduring discussions.

The Hopfield model of neural networks has been thoroughly studied in connection with 
its static and dynamical properties of retrieval, which together determine its usefulness as an associative
memory model. 
Besides its value as a neural model it is possible to look upon the Hopfield model in the broader context 
of complex magnetic systems. In this sense, we can consider it as another kind of long--range spin--glass model
with a different choice of coupling distribution, and 
with the added advantage of having a phase--space structure whose complexity can be controlled.
Both static and dynamical studies of the Hopfield  model have concentrated mainly on the retrieval phase and close to the basin of 
attraction of a stored memory pattern; consequently, in such circumstances, the very rich spin--glass 
structure underlying the free--energy landscape has received little attention so far.

The thermostatics of the Hopfield model has been completely solved assuming that replica symmetry 
holds. While this seems to be the correct solution  within the retrieval zone (except for very low temperatures \cite{amit87}), 
it has not been obvious until now, as far as we know,
which symmetry breaking scheme yields the correct solution within the spin--glass phase of the
model. 

In particular, some corrections, due to symmetry breaking effects, have been found at very low temperatures 
\cite{crisanti}; nonetheless, it has always been sustained that the influence on the retrieval capabilities of the 
network, induced by a breaking in replica symmetry, do not have any noticeable effect on the retrieval performance 
of the system \cite{amit87}. 

The main objective of this paper is twofold. First, to get insight into the underlying structure of the spin--glass
phase of the Hopfield model. This will eventually help to find the correct Ansatz to solve its thermodynamics. 
On the other hand, since it is possible to control the richness of its phase--space structure, ranging from
a simple ferromagnet, when only one pattern is stored, to the SK at the other extreme for a properly taken limit 
of an infinite number of patterns\cite{valee}, it will allow us to extract important
conclusions concerning the influence of such structure on the off--equilibrium dynamics of magnetic systems.

The outline of this paper is as follows: first we describe the model we used and the methods of our simulations. 
Then we present the results and their interpretation within the framework of long--range spin--glasses, and 
finally we discuss our conclusions and suggest further paths of research on the subject.

\section{Model}

The Hopfield model of neural networks is described by the following 
Hamiltonian \cite{hopfield} :
\begin{equation}
H[J]=-\frac{1}{2}\sum_{i j} J_{i j}S_{i}S_{j} \ ,
\end{equation}
where the sum goes over all the possible pairs $i,j$. The variable $S_i$ represents the state of the i--{\em th} 
Ising spin (neuron) at the discrete time $t$, and $J_{i j}$ is the coupling constant between the $i$--{\em th}
and the $j$--{\em th} neurons. The coupling matrix $J$ is built  according to Hebb's rule in order to store a definite set of $p$ 
randomly chosen configurations (patterns):

\begin{equation}
 J_{ij}= \left \{ \begin{array}{ll}
                        \frac{1}{N} \sum_{\mu=1}^{p} \xi_i^{\mu}\xi_j^{\mu}   & \mbox{ if $i \neq j$} \\
                                0                                                                                & \mbox{ if $i = j \; ,$}
                                \end{array}
                        \right. 
\end{equation}

where we denote each of the $p$ patterns as $\vec{\xi}^{\mu}=\{ \xi_i^{\mu} \}$ ($\mu=1 \dots p, i=1 \ldots N$, where $N$ is 
the total number of spins).

The time evolution of the model is governed by a standard heat-bath Monte Carlo process with sequential update, 
and we say that it performs as an associative memory system if each of the $p$ stored patterns is close enough to
an attractor of the dynamics.
In fact, when this happens the structure of the phase space is fairly complex, as a huge number of other 
attractors also appear alongside \cite{amit85}. The simplest of them are the reverse states, i.e., the actual patterns 
with all the spins inverted. Other states which are stable points of the dynamics are the spurious attractors 
(mixture states) and the spin-glass states. A complete solution of the model can be found in \cite{amit87}
and we reproduce in Fig.~1 the phase diagram obtained therein.

The labeled curves delineate the borders of qualitatively differentiated regimes of the Hopfield model in the 
$T-\alpha$ plane, where $\alpha=p/N$. All over the roughly triangular sector enclosed by the $T_M$ line, 
the stored patterns are close to dynamical attractors, therefore the system performs well as an associative 
memory device. Still,  we need to discriminate between two distinct zones associated with the retrieval 
sates which  differ in their phase--space structure. Inside the region below  the $T_c$ line, the retrieval attractors 
represent absolute minima of the free energy. A different structure is found in the region between the $T_c$ 
and the $T_M$ lines where the global minima correspond to spin--glass states totally uncorrelated with the 
patterns. However, the retrieval attractors still stand as local minima of the free energy, thus allowing the system 
to work as an associative memory as long as the dynamics is initiated close enough to a retrieval basin. Then,  
as the $T_M$ line is crossed the retrieval states disappear
abruptly and the spin--glass states become the only ones present. This spin-glass phase covers all the 
region beyond the $T_M$ line and below $T_g$---the latter having a simple mathematical expression, 
namely: $T_g=1+\sqrt{\alpha}$. Finally, above  $T_g$ all the  free--energy 
structure  is lost as the system enters a paramagnetic phase. The line labeled $T_R$ shows the limit of validity of  the replica-symmetry Ansatz.

\begin{figure}
\label{figure1}
\epsfig{file=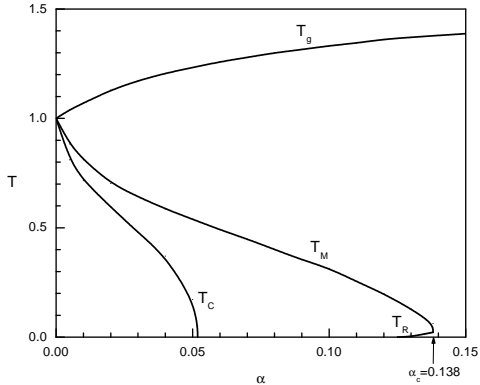,width=80mm}
\caption{Phase diagram for the Hopfield model. See the main text for a brief explanation. After Amit {\em et al} [6]}
\end{figure}

We performed simulations on systems with a number of spins ranging between $128$ and $4000$, with 
particular emphasis in $N=500$ and $N=1000$. Throughout this work the time is measured in units of whole Monte 
Carlo sweeps over the network of $N$ spins.

\section{Results}
In this section we present and describe the results of our simulations for a range of values of the parameter $\alpha$, 
covering  the spin--glass phase of the Hopfield model .

\subsection{Ageing}
In attempting to characterize the dynamics of the model, we first carried out some numerical experiments 
focused on revealing the presence of  slow dynamics in conjunction with history--dependent phenomena, 
i.e., {\em ageing}. These features are most easily found numerically in simulations concerning the two-time 
auto--correlation function $C(t,t')$, which turns out to show a strong explicit dependence on both times 
over a wide range of time scales. For a system that has achieved equilibrium it is expected that $C(t,t')$ is homogeneous in time, depending on $t$ and $t'$ only 
through their difference. However, in spin--glass systems in their glassy phase exhibiting ageing phenomena, 
a much more complicated behaviour arises which reveals the basis for the scenario that has been called 
{\em weak ergodicity breaking} (WEB) \cite{bouchaud}. In the realm of long--range spin--glasses the onset of 
 slow dynamics and history-dependent phenomena below a definite transition temperature, is normally 
associated with
a highly complex free--energy surface with a plethora of metastable states. In a system evolving 
through an energy landscape so complex as that found in long-range spin-glasses, the many basins within 
basins may play the role of dynamical {\em traps} which, having a continuous  distribution of heights without
bounds, can thereby  confine the system in such a way that the average escape--time goes to infinity. 

We can write down the basic features of the WEB scenario more 
explicitly, regarding the 
behaviour observed in the two--time auto--correlation function, as follows:
\begin{equation}
\begin{array}{lr}
\frac{\partial C(t_w,t+t_w)}{\partial t} \leq 0 &  \frac{\partial C(t_w,t+t_w)}{\partial t_w} \geq 0 \\[0.3cm]
\lim_{t\to\infty}\, C(t_w,t+t_w)=0                &  \forall\; \mbox{fixed}\; t_w \, ,
\end{array}
\end{equation}
 where $t_w$ indicates the time that the system has been left to evolve after a sudden quench from infinite
temperature  to a low temperature state, and $t$ is the time measured since the {\em waiting time} $t_w$.

In all our simulations we chose a set of $p$ random patterns, where each of the $\xi_i^{\mu}$'s is taken 
from the following distribution:
\begin{equation}
P(\xi_i^{\mu})=\frac{1}{2}(\delta(\xi_i^{\mu}-1)+\delta(\xi_i^{\mu}+1))
\end{equation}
In order to take measurements of the two--time auto--correlation function we computed numerically the following 
expression:
\begin{equation}
C(t_w,t+t_w)=\Bigl [ \frac{1}{N}\sum_{i=1}^N S_i(t_w)S_i(t+t_w) \Bigr ]_{av} \ ,
\label{web}
\end{equation}
where we have denoted by $[\ldots]_{av}$ an average taken over several realizations of the random patterns and 
thermal baths. The abrupt quench from infinite temperature to the glassy phase was emulated by always 
starting our simulations from random initial configurations.

In many occasions in this paper we shall refer to the SK model for comparison, and consequently we have 
included in Figure~2 the results of a simulation carried out on that model---which although performed 
by us for this work, it is a result  that had already been reported in the literature \cite{yoshino}. This 
figure corresponds to a system with $N=1000$ spins and with couplings taken from a Gaussian distribution 
normalized so that $T_g=1$. In this case the temperature of the thermal bath 
was chosen as $T=0.4T_g$. The graph shows $C(t_w,t+t_w)$ vs. $t$ for different waiting times 
ranging from $t_w=2$ to $t_w=8192$. 

The first result for the Hopfield model is shown in Fig.~3, and it pertains to a system 
with $N=1000$, $\alpha=1$ and $T=0.4T_g$; inside the spin--glass phase and far enough from the  $T_M$ line
to avoid strong size effects. The behaviour of the auto--correlation function confirms clearly the 
presence of ageing in the Hopfield  model in this region, in agreement with 
the WEB picture described by Eqs. (\ref{web}).  The curves show an explicit dependence on both times indicating that 
equilibrium has not been attained within the time of the simulation. Furthermore, its decay becomes slower for 
longer waiting times as it is expected within the context of ageing phenomena. It is worth noticing that the 
qualitative features of the graph are roughly similar to those found in the SK model, in particular for the 
longest waiting times.

\begin{figure}
\label{figure2}
\epsfig{file=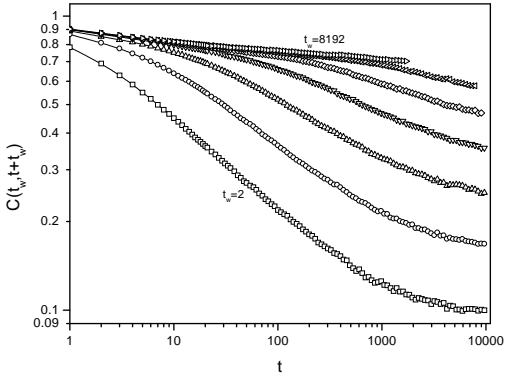,width=80mm}
\caption{Two--time auto--correlation functions $C(t_w,t+t_w)$ vs. $t$ for the SK model, with 
$N=1000$, at $T=0.4T_g$ and with couplings taken from a Gaussian distribution. The waiting 
times are: $2, 8, 32, 128, 512, 2048, 8192$. The exhibited data are an average over $400$ realizations 
of the couplings, initial conditions and thermal baths.}
\end{figure}

\begin{figure}
\label{figure3}
\epsfig{file=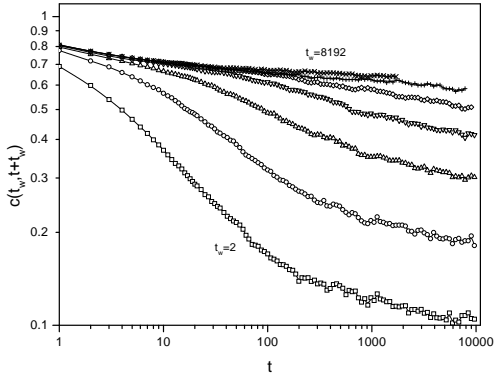,width=80mm}
\caption{Two--time auto--correlation functions $C(t_w,t+t_w)$ vs. $t$ for the Hopfield model, with $N=1000$, 
at $T=0.4T_g$ and $\alpha=1$. The waiting times are: $2, 8, 32, 128, 512, 2048, 8192$. The exhibited data are  
an average over 200 realizations of the set of memory patterns, initial conditions and thermal baths.}
\end{figure}

From the standpoint of this result it is expected that similar features 
of ageing are found for larger values of $\alpha$,
since we know that for $\alpha \to \infty$ the Hopfield model converges 
to the SK model.  A more 
delicate point to be concerned with refers to the
 behaviour of the Hopfield model in regions of its spin--glass 
phase  closer to the retrieval zone. To address that issue we 
studied the dynamics of the model for a value of $\alpha=0.2$, which is 
close to the retrieval zone and at the same time lies 
at a safe distance from the critical $T_M$ transition line, 
whose exact location shifts towards larger values of $\alpha$ due to
finiteness of the system. Besides, we had to deal with 
other manifestation of finite size effects, in this case related to the less
degreee of 
frustration in the couplings as the value of $\alpha$ is lowered. 
It is worth stressing that this strong dependence on $N$ was not
observed for larger values of $\alpha$ (see below). This can
be easily understood in the following terms: higher values of $\alpha$
imply more frustration in the couplings, and so higher energy barriers 
between metastable states. For small values of $\alpha$ the system
can then quickly thermalize, and the time it requires for thermalizing
 depends strongly on $N$. On the other hand, for large values of $\alpha$
the sizes of the energy barriers make the thermalizing time increase far
beyond our simulation times and therefore we observe non--equilibrium 
phenomena that do not  depend on the size of the system.

To analyse the finite size effects, we carried out simulations with 
system sizes ranging from $N=500$ to $N=4000$. 
In Fig. 4 we show $C(t_w,t+t_w)$ vs. $t$ for $t_w=128$ 
and $N=500, 1000, 2000, 4000$ (hollow symbols 
from top downwards), where we see that even for the 
largest size we considered the dependence on $N$ is 
still strong. Since larger systems are beyond our 
computational capacity we decided  to extrapolate the 
data to $N\to\infty$, assuming that the 
$C_N(t_w,t+t_w)$ varies smoothly with $1/N$, and keeping 
terms up to second order.  The lowest curve (full symbols) 
in Fig. 4 corresponds to the extrapolated values obtained 
for $t_w=128$. In Fig. 5 we plot all the extrapolated
curves $C(t_w,t+t_w)$ vs. $t$. 

\begin{figure}
\label{figure4}
\epsfig{file=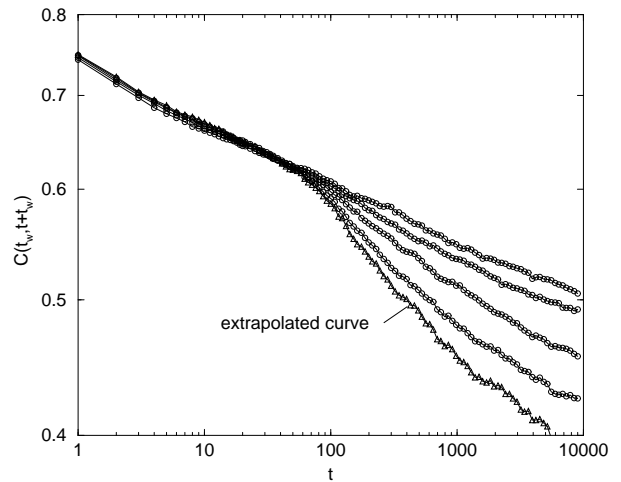,width=80mm}
\caption{Two--time auto--correlation functions $C(t_w,t+t_w)$ vs. $t$ for the Hopfield model, with $t_w=128$, 
at $T=0.4T_g$,  $\alpha=0.2$ and different sizes ($N=500$, $1000$, $2000$ and $4000$ from top downwards). The
lowest curve corresponds to the extrapolated values.}
\end{figure}

\begin{figure}
\label{figure5}
\epsfig{file=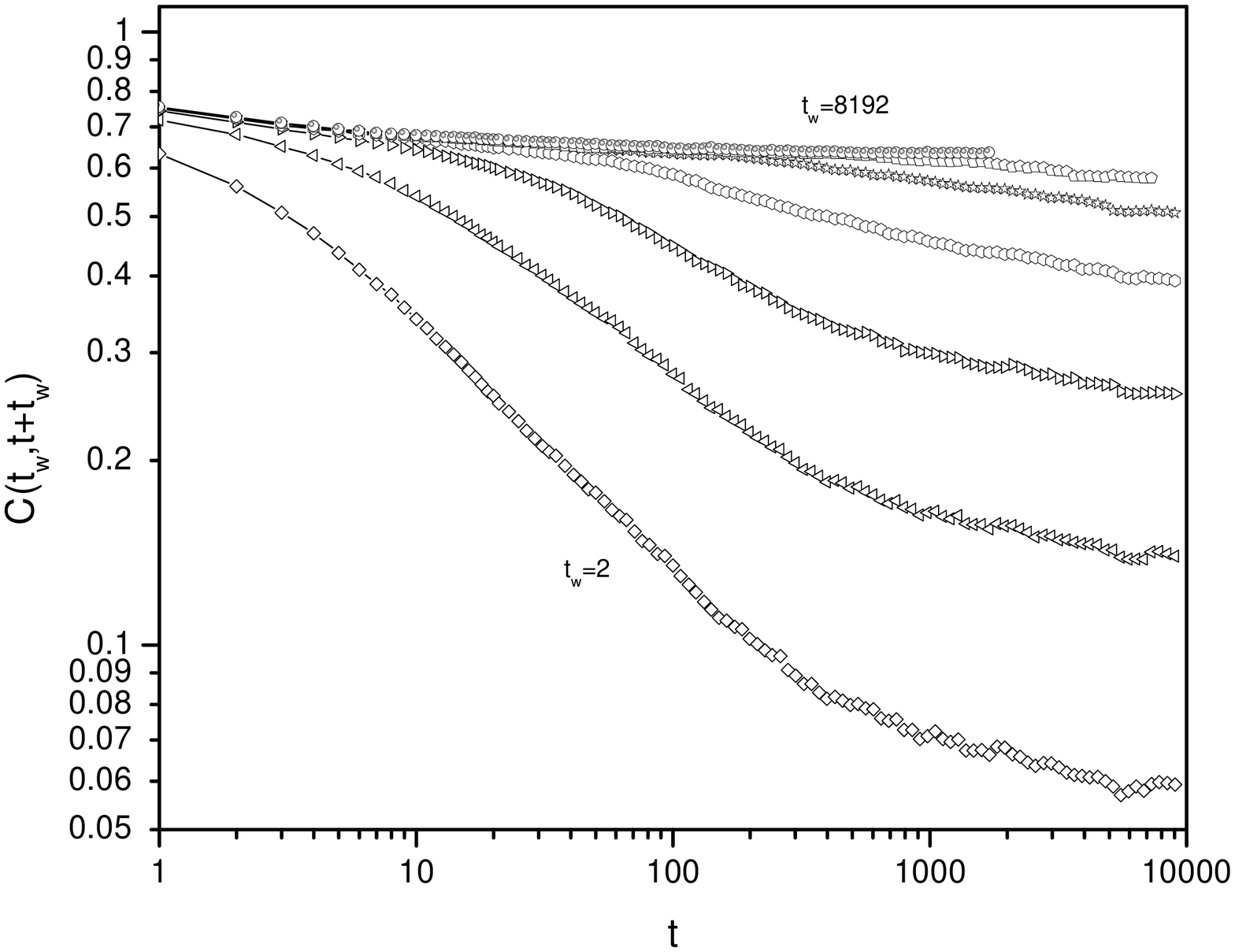,width=80mm}
\caption{Extrapolated two--time auto--correlation functions $C(t_w,t+t_w)$ vs. $t$ for the Hopfield model,
at $T=0.4T_g$ and $\alpha=0.2$. The rest of the parameters are the same as those of Fig.~2}
\end{figure}

In Figs. 6 and 7 we show the results for $\alpha=10$ and $\alpha=100$ 
(with $N=1000$ and $N=500$ respectively, since as 
$\alpha$ increases the results have less dependence on $N$).
Let us observe that the behaviour shown by the Hopfield model
in this range of alpha values is much alike that of the
SK model.

\begin{figure}
\label{figure6}
\epsfig{file=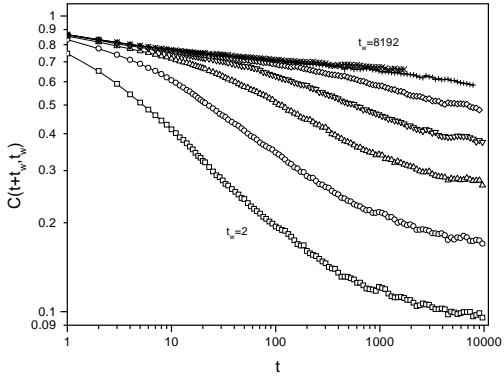,width=80mm}
\caption{Two--time auto--correlation functions $C(t_w,t+t_w)$ vs. $t$ for the Hopfield model, with $N=1000$, 
at $T=0.4T_g$ and $\alpha=10$. The rest of the parameters are the same as those of Fig.~2}
\end{figure}

\begin{figure}
\label{figure7}
\epsfig{file=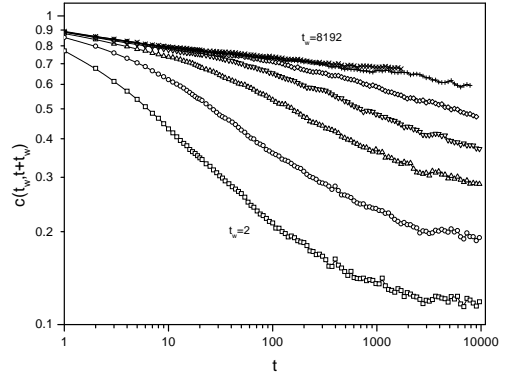,width=80mm}
\caption{Two--times auto--correlation functions $C(t_w,t+t_w)$ vs. $t$ for the Hopfield model, with $N=500$, 
at $T=0.4T_g$ and $\alpha=100$. The rest of the parameters are the same as those of Fig.~2}
\end{figure}

Another quite relevant issue is finding an appropriate scaling law for the ageing curves. 
As has already been pointed out in the literature \cite{cugliandolo2,parisi}, systems that exhibit a continuous 
distribution of relaxation times, like the SK, may not obey a simple scaling relation. 
Instead, for systems with a unique {\em relevant time--scale} (for instance, the p--spin model) it is possible to 
find simple laws that cause all the ageing curves to collapse into a single one. As an example, let us put forward 
the simplest of those scalings, usually called {\em naive scaling}, which takes the following form \cite{bray}:
\begin{equation}
\label{naive}  
C(t_w,t+t_w)=f\Bigr ( \frac{t}{t_w} \Bigl ) \, .
\end{equation}
A likewise dependence is observed in other models that can be well described by theories in which 
full replica symmetry holds \cite{bray}.
The graph in Fig.~8 shows $C(t_w,t+t_w)$ vs. $t/t_w$ for the data displayed in Fig.~3 ($\alpha=1$), and 
brings the Hopfield model even closer to the SK, as it depicts exactly the same patterns found in that 
model\cite{parisi}. That is, on the one hand the roughly common crossing--point at $t/t_w$ (with a corresponding value of 
approximately $1-T/T_g$ on the auto--correlation axis); on the other, the dependence shown by $C(t_w,t+t_w)$ 
for increasing values of $t_w$ when $\tau\equiv t/t_w$ is kept fixed: it increases when $\tau>1$, whereas it decreases 
for $\tau<1$.

\begin{figure}
\label{figure8}
\epsfig{file=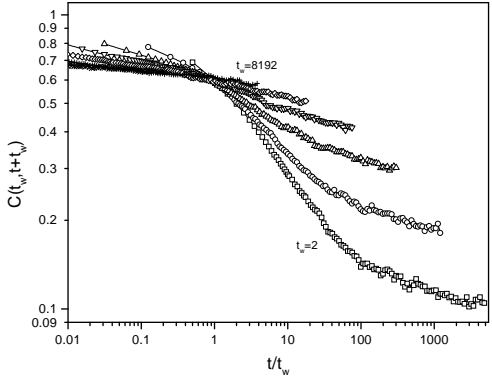,width=80mm}
\caption{Two--time auto-correlation functions $C(t_w,t+t_w)$ vs. $t/t_w$ for the Hopfield model, with $N=1000$, 
at $T=0.4T_g$ and $\alpha=1$. The rest of the parameters are the same as those of Fig.~2}
\end{figure}

The observed lack of agreement between the scaled curves serves as an indication of the complexity 
of the many time scales involved---as was already mentioned in connection with the SK model. 
Next we tried the following more complex expression, which had already been used on the SK model 
yielding a better scaling \cite{parisi}: 
\begin{equation}
\label{log}
C(t_w, t+t_w)=f\Bigl (\frac{ln(t+t_w)}{ln(t_w)} \Bigr ) \, .
\end{equation}
In Figs.~9--11 we have plotted $C(t_w, t+t_w)$ vs. $ln(t+t_w)/ln(t_w)$
for $\alpha=0.2, 1$ and $10$, respectively (the curves for $\alpha=0.2$
where obtained from the extrapolated values displayed in Fig. 5).
By analysing the graphs,
we notice that the data fall into two different groups,
depending on the particular age (waiting time) of the system.
In the first place, the curves for  $t_w=2$ and $t_w=8$, which are
associated to a very young system, do not show a good scaling 
in any of the figures. On the other hand, the data corresponding to
all the other (longer) waitings times are appreciably better scaled
by relation~\ref{log}. As regards Fig.~11 it is worth noticing that the
graph bears a remarkable resemblance to the one obtained for the SK model
in reference \cite{parisi} using the same scaling expression \ref{log}.
This also applies to the  little departures from a correct superposition 
observed when $t>>1$.
Thus, we can conclude that in the spin--glass phase we do not find
any critical value of $\alpha$ (at least larger than $0.2$) below
which the replica symmetry Ansatz holds. On the contrary, our
evidence suggests that the spin--glass phase has a phase--space
structure very similar to that observed in the Sherrington--Kirkpatrick
model.

\begin{figure}
\label{figure9}
\epsfig{file=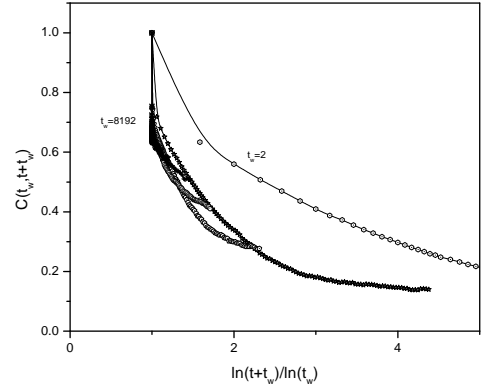,width=80mm}
\caption{Extrapolated two--time auto-correlation functions 
$C(t_w,t+t_w)$ vs. $\ln(t+t_w)/\ln(t_w)$ for the Hopfield 
model, at $T=0.4T_g$ and $\alpha=0.2$. 
The rest of the parameters are the same as those of Fig.~2}
\end{figure} 

\begin{figure}
\label{figure10}
\epsfig{file=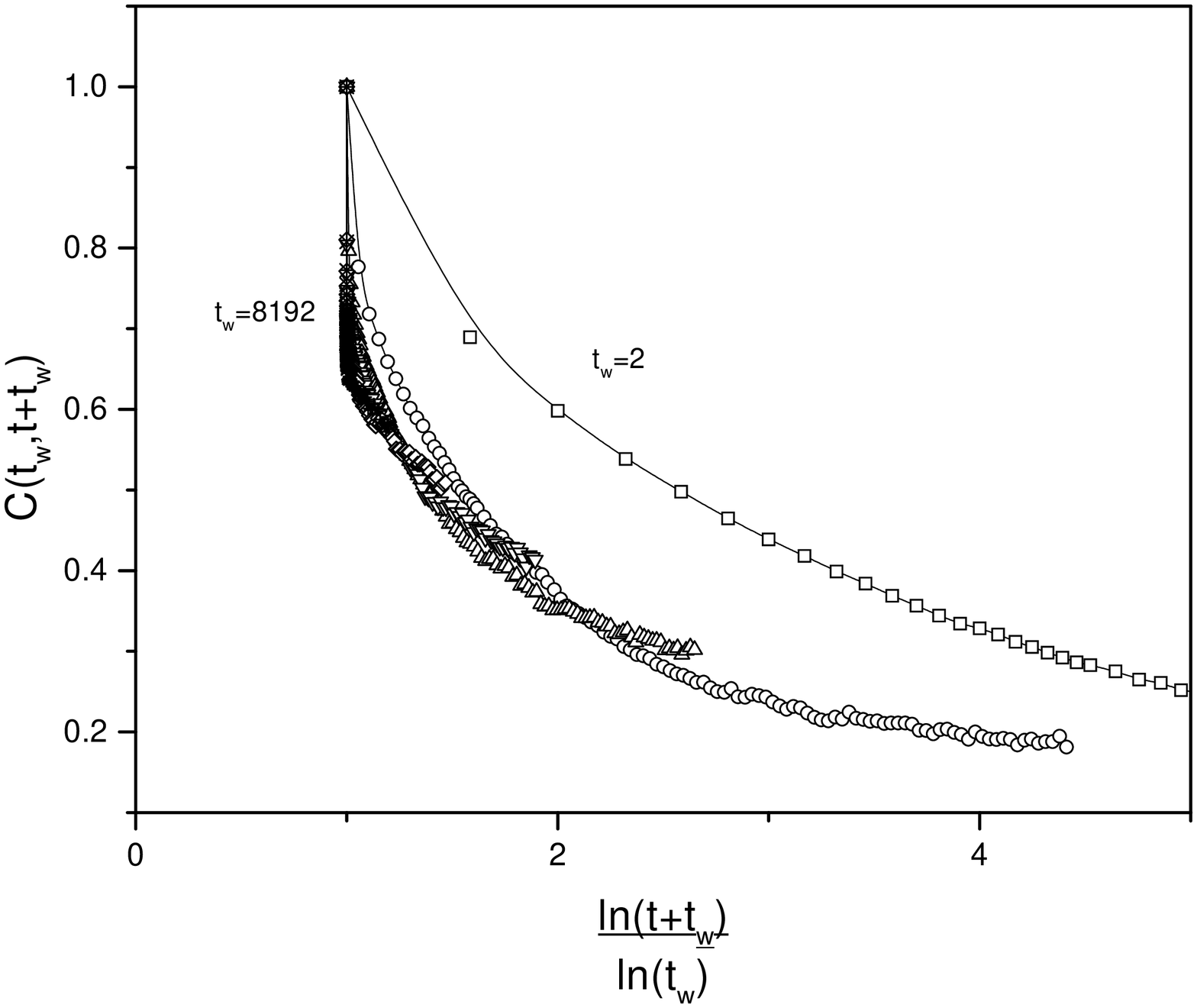,width=80mm}
\caption{Two--time auto-correlation functions $C(t_w,t+t_w)$ vs. $\ln(t+t_w)/\ln(t_w)$ for the Hopfield model, with $N=1000$, 
at $T=0.4T_g$ and $\alpha=1$. The rest of the parameters are the same as those of Fig.~2}
\end{figure}

\begin{figure}
\label{figure11}
\epsfig{file=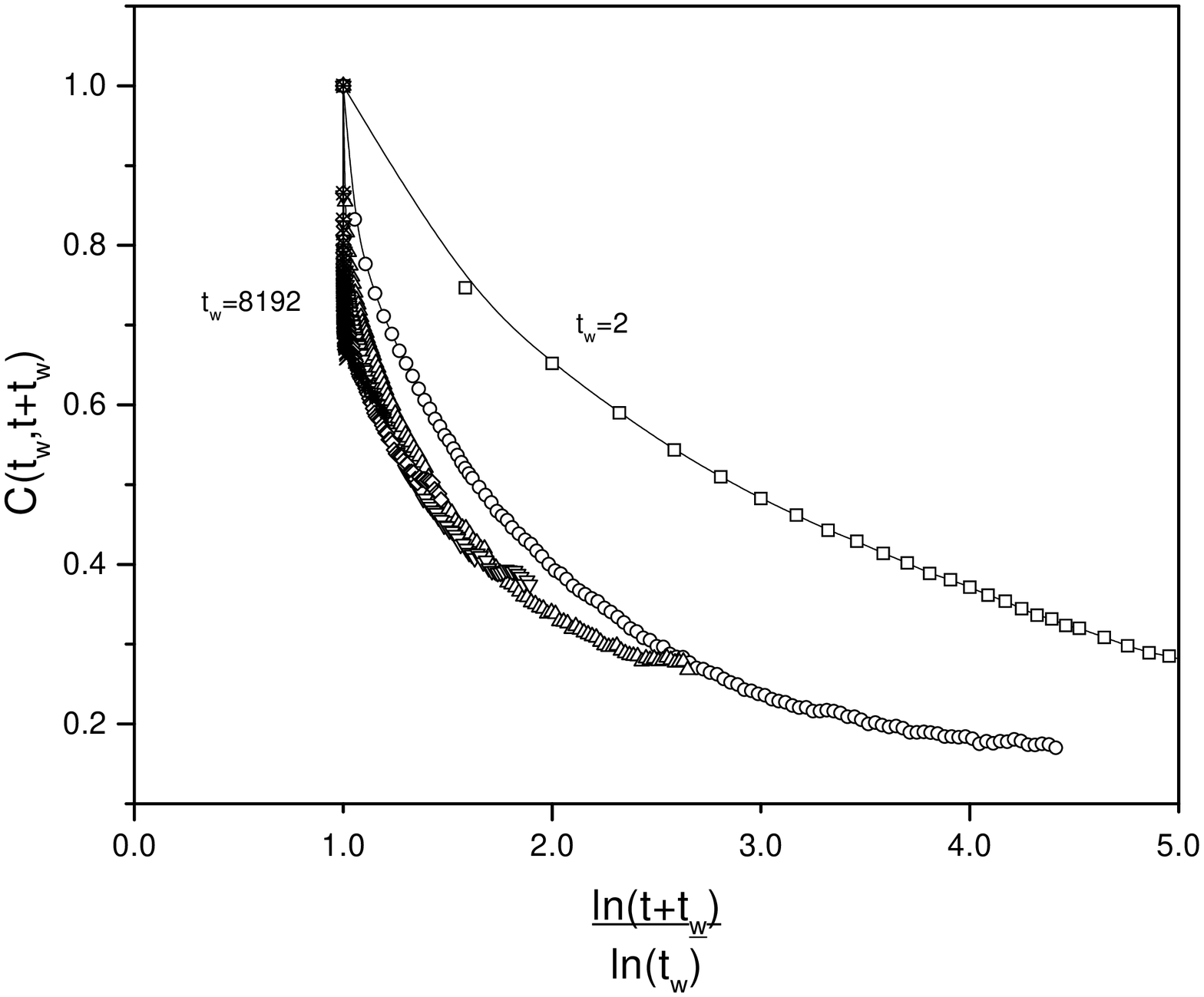,width=80mm}
\caption{Two--time auto-correlation functions $C(t_w,t+t_w)$ vs. $\ln(t+t_w)/\ln(t_w)$ for the Hopfield model, with $N=1000$, 
at $T=0.4T_g$ and $\alpha=10$. The rest of the parameters are the same as those of Fig.~2}
\end{figure}

\subsection{Decay of the Magnetization}

In systems that possess a huge number of metastable states in the presence of quenched disorder, like the SK, the decay 
to equilibrium of physical quantities usually obeys power laws. 
We ran a series of simulations in which we started the Hopfield model from a {\em totally magnetized} state, i.e. $S_i(0)=1 \, \forall i$, and let it relax in contact with a thermal bath and no external fields applied. In this case we monitored the quantity :
%In particular, the decay of the magnetization in the SK 
%model has been studied via numerical simulations in reference \cite{parisi} for different lattice sizes. The %precise measure 
%in which the Hopfield model shares this features may help to clarify how much alike their respective phase 
%spaces are. Indeed, a close qualitative agreement in the intermediate zone on the $\alpha$--range may hint %once more at an intrinsic connection between these to models.
\begin{equation}
m(t)=\Bigl [\frac{1}{N}\sum_i S_i(t) \Bigr ]_{av} \, .
\end{equation}

Then we fitted the obtained data with the same expression used in \cite{parisi}:
\begin{equation}
m(t)=m_{\infty}+\tilde{m}t^{-\delta(T)}
\end{equation}

The results of the simulations are presented in Fig.~12, where we have plotted $\delta(T)$ vs. $T$ for a system with $N=500$ and 
$\alpha=1$. The exponent turns out to be a linear function of temperature in a striking agreement with the behaviour shown by the 
SK model \cite{parisi}.

\begin{figure}
\label{figure12}
\epsfig{file=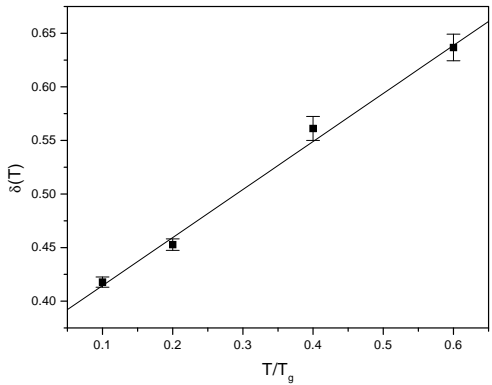,width=80mm}
\caption{Magnetization-decay exponent $\delta(T)$ vs. $T/T_g$. The data corresponds to a system with $N=500$ and $\alpha=1$}
\end{figure}

\subsection{Distribution of Overlaps}

The overlap distribution contains much information regarding the geometric structure of phase space, and its asymptotic 
form may unveil a complex arrangement of many metastable states. We performed a numeric computation of the time--dependent 
overlap--distribution $P(q,t)$. To that purpose we started randomly a set of $15$ replicas of the system and let them evolve with 
independent thermal baths and no couplings between them; at prescribed times we took measurements of all the possible overlaps 
between the replicas, and repeated this procedure for several realizations of the set of random patterns. Finally, we made a 
histogram with all the gathered data.

As it evolves, each system shall probe different regions of the complex free--energy landscape, searching stochastically for 
basins of increasing depth. Hence, their mutual overlaps---at large times---are expected to render some insight into the 
deeper structure of phase space. The corresponding overlaps between each pair of replicas were evaluated according to the following expression:
\begin{equation}
q^{\alpha,\beta}(t)=\frac{1}{N}\sum^N_{i=1}S^{\alpha}_i(t)S^{\beta}_i(t) \ ,
\end{equation}
where the superscripts denote different replicas. 
The results of this simulation can be seen on Fig.~13 
and correspond to a system 
size of $N=128$ spins and $T=0.4Tg$. As we started the 
replicas from random initial conditions, for short times the distributions must 
be approximately Gaussian.
Subsequently, the nearly Gaussian shape 
distorts by broadening at the base, reaching an 
almost uniform distribution for $t=16$ (stars in the graph). From that time on the curve 
begins to develop two clearly defined peaks, which grow in height and sharpen while converging asymptotically to the final shape for 
long times. It is also interesting to notice that the data converge rapidly to a limiting value for $q=0$ which is different from zero. 
From the above mentioned facts we conclude that the distribution for $t\to\infty$ consists of two Dirac's deltas (corresponding 
to $\pm q_{EA}$) plus a continuous distributions between them. This is the same result that has been obtained both numerically 
\cite{yoshino} and analytically for the SK model.

The Edwards--Anderson order parameter stands a geometric indicator that roughly establishes the {\em size} of the deepest basins 
visited by the system in its evolution. The usual way of estimating its value numerically is by measuring the position of the two 
peaks in the distribution $P(q,t)$ for $t\to\infty$. Hence, from the data of our simulations we obtain approximately
$q_{EA}^{(Hop)}\approx 0.73$. Furthermore, we have tested this result against the value predicted by the replica--symmetric 
solution of the Hopfield model by solving numerically the saddle--point equations of\cite{amit87}; the calculations 
yielded $\bar{q}_{EA}^{(Hop)}=0.6167$, which differs considerably from the measured one and in fact falls quite close 
to the value obtained for the SK model within the replica symmetric approximation \cite{sherrington}. These results may 
be compared with the value of the order parameter predicted by Parisi's solution of the SK model, which is given with 
good approximation by the following expression \cite{binder}:

\begin{equation}
q_{EA}^{(SK)}=1-\frac{3}{2}(\frac{T}{T_g})^2 \ .
\label{qsk}
\end{equation}

For $T/T_g=0.4$ equation \ref{qsk} yields $q_{EA}^{(SK)}=0.76$, in good agreement with our data for the Hopfield model. The slight discrepancy  might be due to  an $\alpha$-dependence in the Hopfield model jointly with size effects.

These facts altogether support strongly what has already been suggested by our ageing simulations on the Hopfield model---that the dynamics for long times is essentially that of the SK model.

\begin{figure}
\label{figure13}
\epsfig{file=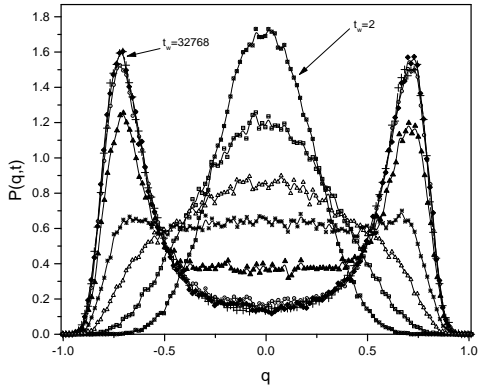,width=80mm}
\caption{Overlap distribution $P(q,t)$ vs. q, for a system with $N=128$ and $\alpha=1$. The different curves correspond to the following times:$2, 4, 8, 16, 64, 512, 4096, 32768$. A total of $15$ replicas were used in each of $744$ runs with different realizations of the random patterns and thermal noise.}
\end{figure}

\section{Conclusions}
In our simulations we have analysed the glassy dynamics of the Hopfield model in the intermediate region between its 
retrieval phase and the SK-limit for $\alpha\to\infty$. So far, no analytical  solution has been found for the model's 
dynamics that incorporates a full breaking of replica symmetry; therefore, our results may shed some light in the 
asymptotic behaviour that ought to be required in analytical treatments of the problem.

The first and foremost conclusion is that the Hopfield model in its spin--glass phase---for intermediate values 
of $\alpha$---shows a complex dynamics which fits neatly into the scenario of {\em weak ergodicity breaking}. 
Comparing the two--time auto--correlation curves found for the Hopfield and SK models, we conclude that 
for long times both systems behave similarly in the whole range of $\alpha$'s we studied. The numerical evidence 
we collected supports the conclusion that the phase--space structure of the Hopfield model is much more 
complex than that expected for a long--range magnetic system in  which replica symmetry holds. Likewise, the 
failing in trying to collapse all the ageing curves into a single one by means of a simple scaling law, may as 
well indicate the presence of a continuous range of temporal scales in the dynamics.

In the intermediate region on the $\alpha$ scale the decay of the magnetization was found to obey a power law, 
indicating the existence of a great number of metastable states. In addition, the decay exponent showed a 
linear dependence on temperature, notably alike the observed in the SK model. This last fact let us go a step 
further and suggest a very similar organization of the vast collection of metastable states in the two models.

The characteristics of the overlap distribution at long times contribute some important facts regarding the 
structure of the most profound basins in the free--energy hyper--surface. On one side we noticed the 
evidence of a continuous portion in the distribution, and on the other, the Edwards--Anderson order 
parameter that we obtained for the Hopfield model---for $\alpha=1$ and $T=0.4T_g$---is very much 
close to the theoretical value found analytically from Parisi's theory for the SK model. Once more, these 
facts point at an inherent resemblance of the two models under comparison.

Summing up the numerical results altogether, we have arrived at the following picture for Hopfield model in 
the studied range of parameters: {\em in the spin-glass phase the free-energy landscape of the Hopfield 
model differs from that of the SK model mainly in the regions farthest away from the deepest basins, which 
are precisely the ones that a randomly started configuration is very much likely to visit first. Nevertheless, 
as the system evolves and diffuses into the rugged  high--dimensional free--energy surface it encounters 
a geometric structure that is essentially the same as for the SK model}.

Among the extensions of the present work we mention that there 
are ongoing studies aimed at determining 
the nature of the $\alpha$--dependence of the phase--space structure 
in the retrieval phase below the $T_M$ 
transition line,
where the 
simultaneous presence of different types of attractors may lead 
to the emergence of complex ageing behaviour.

This work was partially supported by grants from Consejo Nacional de
Investigaciones Cient\'{\i}ficas y T\'ecnicas CONICET (Argentina),
Consejo Provincial de Investigaciones Cient\'{\i}ficas y Tecnol\'ogicas
CONICOR (C\'ordoba, Argentina) and Secretar\'{\i}a de Ciencia y
Tecnolog\'{\i}a de la Universidad Nacional de C\'ordoba SECYT (C\'ordoba,
Argentina). D.A.S. was partially supported by CNPq (Brazil).

\end{multicols}

\end{document}